\documentstyle[epsfig]{aipproc}

\begin{document}
\title{Cosmic $\gamma$-Ray Bursts as a Probe of  \\ 
Star Formation History}

\author{Enrico Ramirez-Ruiz$^{*,\dagger}$, Edward
E. Fenimore$^{\dagger}$ \& Neil Trentham$^*$} 
\address{$^*$Institute of Astronomy, University of Cambridge, Madingley Road,
Cambridge CB3 0HA, United Kingdom.\\
$^{\dagger}$D436, Los Alamos National Laboratory, Los Alamos, New
Mexico 87545, U.S.A.}

\maketitle

\begin{abstract}
 
The cosmic $\gamma$-ray burst (GRB) formation rate, as derived from
the variability-luminosity relation for long-duration GRBs, is compared
with the cosmic star formation rate. If GRBs are related to the
collapse of massive stars, one expects the GRB rate to be approximately
proportional to the star formation rate. 
We found that these two rates have similar slopes at low redshift.   
This suggests that GRBs do indeed track the star formation rate of the
Universe, which 
in turn implies  that the formation rate of massive stars
that produce GRBs is proportional
to the total star formation rate.
It also implies that we might be able to use GRBs as a probe of the cosmic 
star formation rate at high redshift. 
We find that the cosmic star formation rate inferred from the
variability-luminosity relation increases 
steeply with redshift at $z>3.0$.  This is
in apparent contrast to what is derived from measurements of the
cosmic star formation rate at high redshift from optical observations
of field galaxies, suggesting
that much high-$z$ star formation is being missed in the optical 
surveys, even after corrections for dust extinction have been
made.  

\end{abstract}

\section*{The Cosmic Star Formation Rate}

The variation of the total cosmic star formation rate with
redshift $z$ $-$ the SFR plot $-$ is conventionally determined by
measurements of the H$\alpha$ luminosity of galaxies at $z < 0.2$ and
of UV luminosities at $z>0.2$ (Madau et al.~1996).   

Two major complications in constructing the SFR plot at any $z$ are (i) the
need to correct UV luminosities for dust extinction, and (ii) the need to
assume a stellar IMF.  However, both of these seem to have been addressed
with considerable success: (i) by adopting the corrections of Calzetti 
(1997), which when applied to the CFRS data of Lilly et al.~(1996) 
produce a SFR plot similar to that generated from infrared {\it ISO}
observations (Flores et al.~1999), which are
sensitive to the absorbed and re-radiated UV flux, and (ii) by
using an IMF that flattens below 1 M$_{\odot}$ e.g.~that of Kroupa et
al.~1993, which seems to be universal
(Gilmore \& Howell 1998).  The SFR plot constructed with
both of these assumptions, when integrated over cosmic time, reproduces the 
correct local stellar density of $\Omega_* \sim 0.005$ that we derive from a
number of methods.  We can therefore have considerable confidence in
current determinations of the SFR plot at $z < 3$ (Somerville et
al.~2000). 

At $z>3$, there are additional, potentially more serious, complications.
The only types of high-$z$ galaxies whose contributions to the SFR plot have
been unambiguously determined are the Lyman-break galaxies (LBGs; Steidel
et al.~1999).  Lyman-$\alpha$ (Ly$\alpha$) emitters may contribute
too, but current indications 
are that their contribution is small (10$^{-2}$ M$_{\odot}$ yr$^{-1}$
Mpc$^{-3}$ compared to 10$^{-1}$ M$_{\odot}$ yr$^{-1}$ Mpc$^{-3}$ for
the LBGs; Hu et al.~1998). However, these
Ly$\alpha$-emitters and other galaxies which
might exist but cannot be found by the selection techniques used to find
either Ly-break or Ly$\alpha$ galaxies, 
could have their contributions systematically underestimated.
Such galaxies
might be lost due to $(1+z)^{-4}$ surface-brightness dimming
(Lanzetta et al.~2000), or the
detectability of a small fraction of the members of a population
might be enhanced by 
supernovae that happened to go off in those members 
causing us to miss the bulk of the
sources representing that population 
(the supernovae are not dimmed by
$(1+z)^{-4}$ and can affect detectability if 
$10 \log (1+z) > {\rm mag}_{\rm SN} - {\rm mag}_{\rm gal}$).  
Therefore, indirect constraints are important at such high $z$.  The
constraint from $\Omega_*$ is weak since the contribution to the total
integral of the 
SFR plot is small at high $z$ since ${\rm d}t / {\rm d}z$ is small there.
Another constraint is that all the star formation at 
$z>3$ must produce enough metals to enrich the Ly$\alpha$ forest at
$z=3$ (Cowie \& Songaila 1998), but this is also of limited use here, since it
is unclear what fraction of metals escape from the galaxies in which
the stars form.   
A potentially more powerful probe is given by the $\gamma$-ray burst (GRB) 
formation rate plot, which must track the SFR plot closely if the majority of
GRBs originate from the collapse of massive stars. This is what we
investigate here.   
 
\section*{The $\gamma$-Ray Burst Rate}
GRBs are detectable out to the farthest reaches of the 
observable Universe, and
provide information about processes occurring at all cosmic epochs. 
Recent observations suggest that the long-duration GRBs and their
afterglows are produced by highly relativistic jets emitted in
core-collapse supernova explosions. Hence the
redshift distributions of GRBs should track the cosmic star formation
rate of massive stars accurately (Lamb \& Reichart 2000; Blain \&
Natarajan 2000). At present, however, there are too few redshift
measurements with which to estimate the global GRB formation
rate. Nonetheless, these few measured redshifts can be used to
calibrate properties of GRBs that might let them serve as standard
candles. Fenimore \& Ramirez-Ruiz (2000) have suggested that the
spikiness of the burst time 
structure is correlated with luminosity, with smooth bursts being
intrinsically less luminous. In principle, the measured
spikiness combined with the observed flux can be used to obtain
distances much like Cepheid observations give distance estimates from the
pulsation period. Using a sample of 220 bright {\it BATSE} bursts for which
high-resolution light curves were available, Fenimore
\& Ramirez-Ruiz (2000) estimated the evolution of the GRB formation
rate from parameters measured solely at $\gamma$-ray energies.

\begin{figure}[th] 
\centerline{\psfig{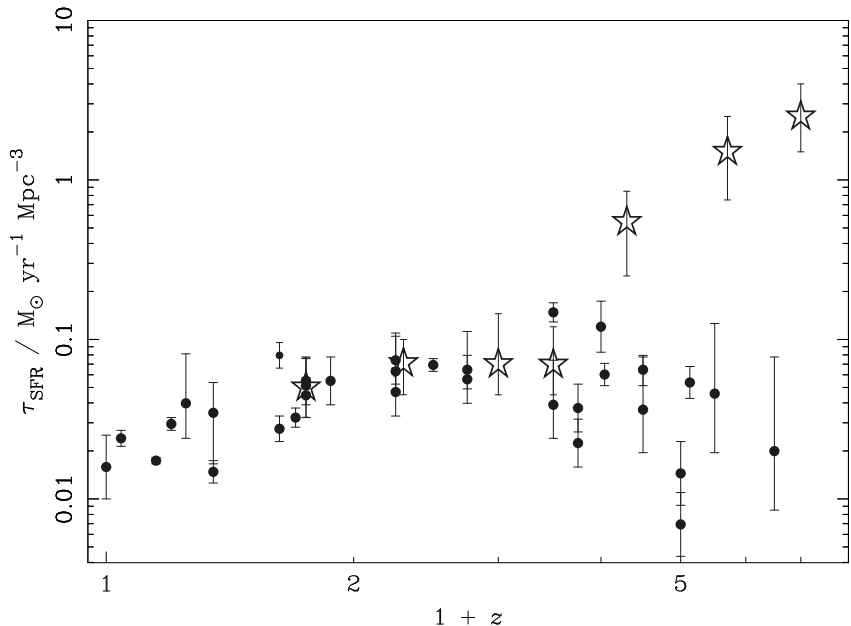}}
\vspace{10pt}
\caption{A summary of the current state of knowledge of the star
formation history of the Universe. The data points plotted are
described in Somerville et al. (2000). The open stars  are the
GRB formation rate obtained by
Fenimore \& Ramirez-Ruiz (2000) normalized to the SFR at $z$=1
($\Omega_{\Lambda}=0.7,\;\Omega_{m}=0.3,\;H_o=\;65\;{\rm km\;s}^{-1}\;{\rm
Mpc}^{-1}$). 
The error bars represent
the systematic uncertainty in the burst formation rate calculated from the
uncertainty in the variability-luminosity relationship, and are
clearly larger than the statistical uncertainty on each data point.}
\label{fig1}
\end{figure}

\section*{The SFR Plot at { \it \lowercase{z}} $<$ 3.0}
At low redshift, the GRB formation rate (BFR) plot tracks the SFR plot
in shape (see Figure 1).  
This is the redshift regime where we have considerable
confidence in the SFR plot (see $\S$ 1).  This concordance suggests: 
\vskip 1pt
\noindent
(i) the idea that GRBs do indeed track the SFR of the Universe.
This would imply that the formation rate of massive stars 
that produce GRBs is proportional
to the total SFR, as would follow if the IMF does not
vary substantially over this $z$ range; 
\vskip 1pt
\noindent 
(ii) that we can use GRBs as a high-$z$ probe.  
They are not affected by extinction or surface-brightness dimming.
So one might therefore
expect the high-$z$ BFR plot to track the high-$z$ SFR plot.
 
\section*{The SFR plot at  { \it \lowercase{z}} $>$ 3.0}
The high-$z$ BFR plot has a steep slope.  This is different from the
much flatter slope
of the SFR plot as derived from LBG observations (Steidel et
al.~1999). If the BFR plot does indeed track
the SFR plot at high-$z$ (as it does at low-$z$), then this would suggest
that much high-$z$ star formation is being missed in the optical 
surveys, even after corrections for dust extinction have been
made.  

Interestingly, Lanzetta et al.~(2000) find evidence for a very
steep SFR plot based on an analysis of high-$z$ galaxies in the Hubble
Deep Fields in which they consider selection effects that can arise 
from surface-brightness dimming.  An additional source of 
star-formation at high-$z$ could
come from luminous {\it SCUBA} galaxies (Blain et al.~1999) which could be
missing from optical surveys (e.g.~Smail et al.~1999) because
they are heavily extinguished by dust. 
  
One important caveat is that the redshift range ($z<3.0$) in which
the BFR plot tracks the SFR plot is similar to  
the redshift range in which GRBs were used to calculate the
variability-luminosity 
relation (there is only one event with $z >$ 3.0).  The application of
this relation to faster variabilities and therefore higher luminosities (as
is required to derive the BFR plot at $z>3.0$) depends on an
extrapolation of this relation.  If the extrapolation fails, this
weakens our conclusion about the steep slope of the BFR at high-$z$. 
Also, if the fraction of massive stars that
produce GRBs is a strong  
function of metallicity, then this could create a redshift-dependence of
the normalization of the BFR plot relative to the SFR plot, 
which would also weaken the assertion that the SFR plot is steep at high-$z$.

\end{document}